*Shteryo Nozharov* [1]
*Petya Koralova - Nozharova* [2]

# THE CORRELATION: MINIMUM WAGE – UNEMPLOYMENT IN THE CONDITIONS OF TRANSITION TO DIGITAL ECONOMY

## 1. Introduction

The adoption of the Directive on adequate minimum wages in EU in 2022 raised many challenges to the EU member-states (European Commission, 2022). The main challenge, which they need to face is to comply the level of their minimum wages with the average wage levels and median wage levels, which fact is based on the formulation of the Kaitz index. In the EU member-states, which are in a process of incomes convergence with the average values for EU and those which are still not Eurozone member-states, this fact will lead to a pressure on governments to increase the minimum wage levels. Countries, like Bulgaria and Romania could have some difficulties in the process of implementation of the Directive and this could negatively affect the unemployment rate in these member-states.

On next place, the transition to digital economy is increasingly reducing the productiveness of low-skilled labor. In this way, the administrative increase of the minimum wage levels in accordance with the provisions of the Directive will force employers to replace low-skilled personnel with machines and software.

Balsmeier and Woerter (2019) focus their research on the attitude of private companies to make investments in digitalization of their economic activities and to evaluate how these processes influence the employment rate. They have done their empirical analysis based on a representative survey amongst Swiss companies. The authors have concluded that the investments in information and communication technologies deployment in the economic activities of companies, have positive effect on the employment of high-skilled workers and respectively a negative impact on the employment of workers who are low-qualified. In this regard, they present a set of measures that national governments need to consider when preparing their social policies. First, the competitiveness of technologically advanced countries will depend on the number of high-skilled workers employed, especially in the production sectors of the national economies. This in turns, will require a flexible labor market, which will easily adapt to the achievements of the science and technology. Second, the life-long improvement of professional skills of workers is of utmost significance so as to be achieved a synchronization between the required qualification for occupying the relevant profession and the knowledge and skills that workers need to have.

[1] *Shteryo Nozharov, chief assistant professor, PhD, University of National and World Economy, Department of Economics, e-mail:* nozharov@unwe.bg
[2] *Petya Koralova-Nozharova, chief assistant professor, PhD, Economic Research Institute at Bulgarian Academy of Science, e-mail:* p.koralova@iki.bas.bg

Another team of authors (Borsova, et. al, 2020) examines two hypothesis, concerning the influence of digitalization on the labor market development. According to the first hypothesis, a clear negative correlation between economic processes digitalization and increase of unemployment rate has not been established yet. They believe that this process will be smoothly happening in the medium-run and the long-run, and the labor market will adapt to the emergence of brand-new professions and to the disappearance of existing ones. In accordance with the second hypothesis, the authors have studied how susceptible are jobs to computerization. In this regard, they have proved that there is a real risk of approximately 50% of jobs in the USA to be replaced with information and communication technologies (computers, algorithms, and robots) in the next two decades.

The publication of Szabó-Szentgróti, Vagvari and Varga (2021) evaluates the positive and negative effects of Industry 4.0 on the development of the labor market. To the positive aspects of digitalization, they refer increase of labor productiveness and competitiveness of enterprises; higher quality of the products and services provided; low amount of the variable costs etc. The negative aspects of digitalization on the labor market, they relate to the increased rate of the structural unemployment; deepening of inequalities amongst the various groups; lack of qualified personnel; loss of approximately 800 million of jobs till 2030; slowdown in the global economic growth; change in the educational methods and acquisition of professional qualification and skills amongst the population.

The purpose of the present research is to identify the direction and strength of correlation between changes in minimum wage levels and unemployment rate in the context of conflicting findings in the scientific literature. An important research question here is whether the business cycle phases influence this process in the conditions of transition to digital economy. In this regard, the present publication contributes to the development of the scientific literature by providing empirical evidence for the opportunities the provisions of the Directive on adequate minimum wages in EU to be implemented in EU member-states with the lowest households' incomes in the European Union.

## 2. Theoretical background

The Nobel prize for economics 2021 of David Card, Joshua Angrist and Guido Imbens raised more questions than answers. The dominant theoretical understanding of the late $20^{th}$ century that the increase of minimum wage levels leads to decrease of employment rate because of higher production costs was challenged. Card and Krueger (2000) argue that such a negative correlation does not necessarily occur. Several statements deserve attention in their publication:

First, the increase of minimum wage levels manifests its effects over time. This could be done in administrative way in a period of economic upswing, and the administrative act through which the minimum wage is increased, could enter into force in a period of economic recession. Consequently, there is a time lag due to the administrative procedures. This creates difficulties for evaluating the labor market reactions to the changes in the minimum wage levels.

Second, the increase in the minimum wage levels probably leads to increase in the final products prices in the respective economic sector. Thus, the negative effects of this increase are borne not by the workers, but by the consumers. This process is observed even in economic sectors with perfect competition, like fast food restaurants.

There exist publications by other authors who have examined the negative correlation between increase of minimum wage levels and the employment rate.

Chong-Uk-Kim and Gieyoung Lim (2018), have made an empirical analysis of 25 OECD member-states. The authors concluded that companies could hire skilled or unskilled personnel so as to maximize their profit, where setting a higher minimum wage levels leads to a decrease in labor demand but does not lead to changes in the labor supply. According to the empirical data presented, it is obvious that "the modest" increase of minimum wage levels has a limited influence on the employment rate. For example, with a 10% increase in the minimum wage level, the employment rate decreases by 0,7%.

The effects of changes in the minimum wage levels on the regional employment and unemployment are studied in the case of strong market economies, such as Germany. In the publication of Bonin et al. (2019) the effects of introducing statutory minimum wage since 2015 on the employment and unemployment rates at regional labor markets in the short run and the medium run are identified. The authors have summarized that the adoption of such type of minimum wage in Germany leads to a significant reduction in the marginal employment rates during the first years of implementing this reform. According to the authors, there is no evidence that the minimum wage level causes significant decrease in the employment rate.

Alison Wellington (1991) has studied what the effects of minimum wage levels on youth employment are. According to him, approximately 21% of the population in the ages between 20-24 years earn minimum wage in the USA. While studying the correlation between the abovementioned variables, the author has accounted the effects of teenage education without considering the seasonal components, which influence the youth employment rate. In the conclusions, he has summarized that there is no significant evidence that the increase of minimum wage levels will have any strong effect on the youth employment rate.

The issue, concerning the influence of minimum wage levels is studied also by Yih-chyi Chuang (2006). The author identifies the influence of minimum wage levels on the number of employed/unemployed youngsters in accordance with the adoption of new economic policies for minimum wage levels regulation in Taiwan. The author applies an empirical model which takes into consideration not only the influence of minimum wage levels on youth employment rate, but also the factors, that influence the labor supply and demand, as well as the components accounting seasonality and regional differences. For measuring the minimum wage level, Chuang uses Kaitz index. The conclusions of the author are that in the short run the introduction of minimum wage amongst youngsters leads to higher employment rates and does not have a reverse effect on the youth unemployment rate. In the long run, however, the youth employment rate will depend on the higher labor supply by them. In this case, the effect on achieving long-term economic growth will be a negative value, as the accumulation of human capital will be slowed down.

Maloney (2009) has studied the correlation between minimum wage levels amongst youngsters and the unemployment rate in New Zealand. According to him, the participation of labor force in the labor market depends both on the wage levels and possibilities of ensuring employment. The author is searching for the answer to the question "is there any evidence that employers tend to substitute low-paid youngsters for high-paid adults in the conditions of increasing minimum wage levels?" As a result of the empirical analysis, Maloney (2009) has found a direct effect of minimum wage levels on the development of the labor market in New Zealand, where this labor market is fully centralized. Taking into account, the seasonality, the effects of the business cycles and other external factors, the author by means of time series has understand that 10% increase in the minimum wage levels leads to a decrease in the youth employment rate by 3,5%. When considering the lack of sufficient qualifying skills amongst youngsters, this percentage increases to 5,7. Maloney (2009) has also inferred the indirect effects of the changes in the minimum wage levels and according to him teenagers who do not earn the same minimum wage levels as adults, are "close substitutes" to young adults at the labor market. In this way the increase in the minimum wage levels has a reverse effect on the teenager's employment rate.

Gavrel, Lebon and Rebiere (2010) evaluate the short-term and long-term influence of minimum wage levels on the unemployment rate in terms of recruitment and vacancy management policies in companies. As a result of the econometric analysis applied, the authors conclude that the introduction of minimum wage levels favors employment rate, but this not always means effective functioning of the labor market. They summarize that when the workers employment in the low-skilled economic sectors is increased, this will lead to an increase in the overall unemployment rate.

This topic is also studied by Bulgarian authors, such as: Tsanov, Shopov, Beleva, Hristoskov, (2017), Beleva, (2017) and others.

### 3. Study area, data, and methods

The scope of the study is focused on EU member-states, which are in a process of income convergence with the EU average values and those which are not Eurozone member-states yet. The representative survey is done for Bulgaria, which is an EU member-state since 2007.

For the purposes of the present analysis, there will be used the following data: statistical information from international databases such as EUROSTAT and World Bank database. The missing information about the period before Bulgarian membership to the EU, is taken from the database of Bulgarian National Statistical Institute. The available data for this period, which is presented in BGN is converted into USD or euros.

The empirical research is done by applying regression analysis through the statistical software STATA. The chosen 30-years period for the analysis (1991-2021) coincides with the time when Bulgaria is a market economy.

## 4. Results and Discussion

Before running the linear regression, we will first establish a scatterplot of the minimum wage levels versus unemployment rate in order to visualize the relationship between these variables and to check for obvious deviations.

*Figure 1. Scatterplot of the correlation minimum wage levels – unemployment rate for the period 1991-2021*

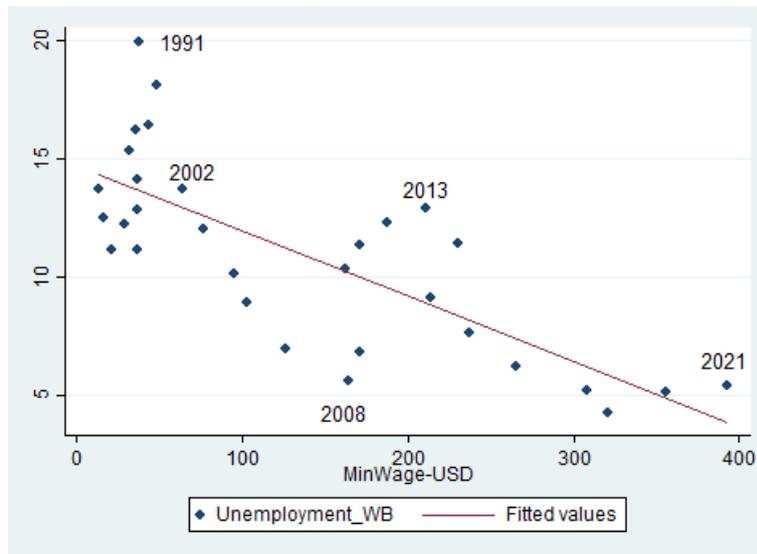

*Source: authors'own calculations based on STATA software and data from the World Bank database, (2022) National Insurance Institute of Republic of Bulgaria (2011,pp.43-44 ), National Insurance Institute of Republic of Bulgaria (2021, pp.30, 40), Country economy (2022), Eurostat (2022)*

The results achieved show that in a 30-year statistical series (1991-2021) – the linear regression has five periods: 1991-1995, 1996-2002, 2003-2007, 2008-2013, 2014-2021. These periods absolutely coincide with the real phases of the business cycles of the Bulgarian economy. It is a well-known fact that the world economic and financial crisis, which started in 2007-2008 affected negatively also the Bulgarian economy, which fact is also reflected in the statistical data, available in the EUROSTAT and World Bank databases.

On figure 1, one can see that in the period 2008-2013, there is a clear reverse trend: the increase in the minimum wage levels has led to an increase in the unemployment rate. The other similar trend, which is observed, is in the time 1996-2002, which is related to a period of crisis for the development of the Bulgarian economy.

Consequently, in a phase of recession or crisis of the business cycle of Bulgarian economy, an increase in the minimum wage levels leads to an increase in the unemployment rate of the country.

This fact provokes an interest to the authors to study the correlation between unemployment rate and minimum wage levels for the period 2008-2013. This decline phase of the business cycle is most clearly expressed in the scatterplot.

In this way we could use the Pearson's correlation coefficient to define the linear correlation between these variables. The results show that the correlation between these variables is very strong and positive and there is a proportionate relationship between them.

*Table 1. Pearson's correlation coefficient (for the period 2008-2013)*

```
(obs=6)

             |  CoefUn~l   MinWage
-------------+--------------------
  CoefUnempl |   1.0000
     MinWage |   0.8918    1.0000
```

*Source: authors'own calculations based on STATA software and data from the World Bank database, (2022) National Insurance Institute of Republic of Bulgaria (2011,pp.43-44 ), National Insurance Institute of Republic of Bulgaria  (2021, pp.30, 40), Country economy (2022), Eurostat (2022*

This strong correlation between the variables allows the regression analysis to be proceeded. Let's check the correlation between the variables through the linear regression model for the period 2008-2013, through the STATA software.:

*Table 2. Linear regression: impact of minimum wage levels on unemployment rate for the period 2008-2013 (yearly data)*

```
. regress CoefUnempl MinWage

      Source |       SS       df       MS              Number of obs =       6
-------------+------------------------------           F(  1,    4) =   15.53
       Model |  27.906126     1   27.906126            Prob > F      =  0.0169
    Residual |  7.18615586    4   1.79653897           R-squared     =  0.7952
-------------+------------------------------           Adj R-squared =  0.7440
       Total |  35.0922819    5   7.01845638           Root MSE      =  1.3404

------------------------------------------------------------------------------
  CoefUnempl |      Coef.   Std. Err.      t    P>|t|     [95% Conf. Interval]
-------------+----------------------------------------------------------------
     MinWage |    .071244    .0180766     3.94   0.017     .0210553    .1214327
       _cons |  -8.347174    4.681783    -1.78   0.149    -21.34589    4.651541
------------------------------------------------------------------------------
```

.

The explanatory power of the R2 model is approximately 80% (79,5%).

This means that during the chosen period, 80% of the changes in the unemployment rate could be explained by the changes in the minimum wage levels.

Coef. (MinWage) is +0,071244. This shows that the average changes in the response variable are associated with one-unit increase in the explanatory variable. In this case, each increase in the minimum wage levels is associated to an increase of the unemployment rate with an average value of +0,071244.

This data will be used so as to be formulated a regression equation, which represents the positive correlation between the minimum wage levels and the unemployment rate for the studied period.

The regression equation is the following:

**Unemployment rate = -8.347174 + 0.071244* Minimum wage levels + ε          (1)**

After making the necessary replacements in the regression equation, it could be concluded the following: if the government increases the minimum wage level with 50 BGN and the determination coefficient is taken into consideration, it could be accepted that in a period of economic crisis, the increase in the minimum wage levels will lead to an increase in the unemployment rate by 6%. Since the unemployment rate depends on various factors, here it is important the direction of the correlation between the studied variables, which is a positive one, especially in a period of economic crisis, which is also proved by the derived regression equation.

Now, it is necessary the reliability of the analysis to be checked.:

The Significance F is < 0.05 and equals 0.0169.

RMSE=1.3404. This is the root mean square error. SI= 0.13. These values are acceptable because when we use yearly data the value of SI needs to be under 30%, which is a reliable value as the explanatory power of R2 is over 0.75.

$P>|t|$. As this value is < 0,05 (0.017), we could conclude that there is a statistically significant correlation between the minimum wage levels and the unemployment rate.

Based on the abovementioned, it could be assumed that the model used is reliable.

In order the abovementioned results to be verified, another control analyzes are done:

1. Regression analysis of the correlation minimum wage levels and the absolute value of unemployment for the studied period.

2. Regression analysis of the correlation between average monthly insurance incomes and unemployment rate for the studied period.

3. Regression analysis of the correlation between the absolute volume of employment in the private sector and the minimum wage levels for the studied period.

The control analyzes have proved the credibility of the model.

## 5. Conclusion

As a result of the conducted regression analysis for the evaluation of changes in minimum wage levels on the unemployment rate for a 30-year period (1991 – 2021) for Bulgaria, the following conclusions could be made:

During a phase of economic downturn or recession of the business cycle of the Bulgarian economy, an increase in the minimum wage levels will lead to an increase in the unemployment rate. The regression analysis has an explanatory power of 80% which means that it is credible. This result could be perceived as a contradiction to the findings of Card and Krueger (2000). Probably these contradictions could be explained by the different structures of the USA economy and the national economies of the new EU member-states, like Bulgaria and Romania.

Another conclusion of the analysis is that the adoption of the Directive on adequate minimum wages in EU should not need to be associated with an administrative annual increase in the minimum wage levels, based on the Kaitz index. In this way, the phases of the business cycle will not be considered. The minimum wage levels should only be increased in phases of upswing and high economic growth rates of the business cycle of the national economies. And vice versa, in the phases of downturn and recession of the business cycle, every increase of the minimum wage levels should be limited.

In addition, it is obvious that in the process of transition to digital economy, it is very difficult to be measured the low-skilled workers' productiveness. In this way, the administrative increase of the minimum wage levels, which is not assigned with the business cycle phases, will force employers to replace human resources with machines and software. This fact will seriously affect the implementation of the national income policies by the governments.

**Acknowledgements**

The present study is approbated amongst Bulgarian private companies and Bulgarian mass media through the kind support of Bulgarian Industrial Association (Nozharov, 2021).

**Notes**

The present paper was presented during the International Scientific Conference 2022 "Economic development and policies: Realities and Prospects. Challenges and Risks in the Conditions of Overlapping Crisis" which was held in the period 21-22.11.2022 and was organized by the Economic Research Institute at Bulgarian Academy of Science.

*Shteryo Nozharov*
*Petya Koralova – Nozharova*

# THE CORRELATION: MINIMUM WAGE – UNEMPLOYMENT IN THE CONDITIONS OF TRANSITION TO DIGITAL ECONOMY

*Abstract: The research is done in the context of the upcoming introduction of new European legislation for the first time for regulation of minimum wage at European level. Its purpose is to identify the direction and strength of the correlation amongst changes of minimum wage and unemployment rate in the context of conflicting findings of the scientific literature. There will be used statistical instruments for the purposes of the analysis as it incorporates data for Bulgaria for the period 1991-2021. The significance of the research is related to the transition to digital economy and the necessity for complex transformation of the minimum wage functions in the context of the new socio-economic reality.*
*Keywords: minimum wage, unemployment, digital economy*
*JEL: J31, J23, E24*